\documentclass[12pt,final]{iopart}
\usepackage{iopams}
\usepackage{graphicx}
\usepackage[numbers,sort&compress]{natbib}
\usepackage{hyperref}

\begin{document}
\title{Highly accurate local basis sets for large-scale DFT
  calculations in CONQUEST}
\author{David R. Bowler}
 \address{London Centre for Nanotechnology, University College
  London, 17-19 Gordon St, London, WC1H 0AH, UK}
\address{Department of Physics \& Astronomy, University College London,
Gower St, London, WC1E 6BT, UK}
\address{International Centre for Materials Nanoarchitectonics
   (MANA), National Institute for Materials Science (NIMS), 1-1
   Namiki, Tsukuba, Ibaraki 305-0044, Japan}
\ead{E-mail: david.bowler@ucl.ac.uk},
\author{Jack S. Baker, Jack T. L. Poulton, Shereif Y. Mujahed}
 \address{London Centre for Nanotechnology, University College
  London, 17-19 Gordon St, London, WC1H 0AH, UK}
\author{Jianbo Lin, Sushma Yadav, Zamaan Raza and Tsuyoshi Miyazaki}
\address{International Centre for Materials Nanoarchitectonics
   (MANA), National Institute for Materials Science (NIMS), 1-1
   Namiki, Tsukuba, Ibaraki 305-0044, Japan}
 \begin{abstract}
   Given the widespread use of density functional theory (DFT),
  there is an increasing need for the ability to model large systems
  (beyond 1,000 atoms).  We present a brief overview of the
  large-scale DFT code \textsc{Conquest}, which is capable of
  modelling such large systems, and discuss approaches to
  the generation of consistent, well-converged pseudo-atomic basis 
  sets which will allow such large scale calculations.  We present
  tests of these basis sets for a variety of 
  materials, comparing to fully converged plane wave results using the
  same pseudopotentials and grids.
\end{abstract}


\section{Introduction}
\label{sec:introduction}

Over the last thirty years, density functional theory (DFT) has
emerged as the leading electronic structure modelling technique, used
in fields as diverse as biochemistry and electronic engineering,
alongside physics, chemistry and materials science.  The ability to
model the atomic and electronic structure of molecules, liquids,
nanoparticles and solids has become a key part of the scientific
method.

However, almost all simulations are performed on a very restricted
number of atoms: typically a few hundred, and very rarely beyond a thousand.  There are various
reasons for the restriction, the most often-cited of which is scaling:
the asymptotic scaling with the number of atoms of standard DFT is cubic in time and quadratic
in memory.  While there has been considerable
effort in developing linear scaling approaches to
DFT\cite{Goedecker:1999pv,Bowler:2012zt} which have been demonstrated
to be capable of calculations on millions of
atoms\cite{Bowler:2010uq,Arita:2014qr}, these introduce 
approximations (typically integrating over energy, so that eigenstates
are not easily available, though can be calculated within a
range\cite{Nakata:2017mw}).  There is therefore a significant gap in
the sizes of systems that can be modelled exactly, lying in the range of thousands
to tens of thousands of atoms.

There are numerous examples of simulations that require this
size of system, for instance: realistic doping in semiconductors; biological
molecules with or without explicit water; nanostructures; compounds
with very dilute compositions; and large scale defects such as
dislocations.  There are, of course, many possible approaches to large
scale electronic structure calculations.  Approximate methods, such as
density functional tight binding (DFTB)\cite{Porezag:1995rs,Elstner:1998sv} are effective, but introduce
some empiricism and/or fitting, reduce basis sets and remove explicit calculation of
certain terms.  Full DFT calculations can be made both more efficient and
to scale better in parallel by judicious choice of basis sets.  The
largest example of a full DFT calculation used a real-space
finite-difference approach to perform calculations on 10,000+ atom
systems\cite{Iwata:2010mf} with a demonstration that 100,000 atoms are
possible\cite{Hasegawa:2013lw}.

Localised basis sets are a common choice for efficient, large-scale
DFT approaches.  Using such a basis, it is easy to form sparse
matrices, and to diagonalise the Hamiltonian exactly.  Examples of
these approaches include the pseudo-atomic orbitals found in the
widely-used \textsc{siesta} code\cite{Soler:2002kn} and the Gaussian basis sets in the CP2K
code\cite{VandeVondele:2005hr}.

Our focus in this paper is the large-scale DFT code,
\textsc{Conquest}, which is capable of exact calculations on systems
with several thousand atoms \cite{Nakata:2017mw,Romero-Muniz:2018pr}
and linear scaling calculations on millions of atoms\cite{Bowler:2010uq,Arita:2014qr}.
\textsc{Conquest} uses a basis set of support functions (SFs) to
represent both the Hamiltonian and either the density matrix or the
Kohn-Sham eigenstates.  These support functions can be further
represented by other functions: either a systematic basis of blip
functions\cite{Hernandez:1997ay}; or numerical pseudo-atomic orbitals
(PAOs)\cite{Torralba:2008wm} (though \textsc{Conquest} often uses a 1:1 mapping of PAOs to
SFs).  \textsc{Conquest} can solve for the electronic ground state
using three key methods: linear scaling
DFT\cite{Bowler:2002pt,Bowler:2010uq,Arita:2014ca} which we will not
discuss further here; multi-site support
functions\cite{Rayson:2009vo,Rayson:2010fk,Nakata:2014ev,Nakata:2015ld},
where a small number of support functions are made from the PAOs from
several atoms (hence multi-site), with the Hamiltonian solved by exact
diagonalisation; and the primitive PAO basis set (one PAO is one SF),
also solved using diagonalisation.

PAOs are convenient and easily generated, but are not systematic in
their convergence.  In this paper, we will describe our approach to generating basis sets of
varying sizes, and will demonstrate that it is possible to produce
relatively modest basis sets that closely reproduce converged plane-wave results
using the same pseudopotentials.

\section{Implementation details}
\label{sec:impl-deta}

The \textsc{Conquest} code has been described in detail
elsewhere\cite{Bowler:2001yo,Bowler:2002pt,Miyazaki:2004ee,Arita:2014ca,Nakata:2015ld}; 
nevertheless, some details of the implementation may be useful to
assist the reader, and we summarise these here.

The central part of the \textsc{Conquest} code relates to the creation
of matrices and their use in finding the ground state of the system
being studied.  When using PAOs as the basis set (as in this paper),
we form the overlap matrix elements, and the kinetic and non-local
pseudopotential contributions to the Hamiltonian matrix elements using
the high accuracy approach pioneered by the \textsc{siesta}
code\cite{Soler:2002kn}.  The remaining parts of the Hamiltonian
matrix elements (from the local part of the pseudopotential as well as
the Hartree and exchange-correlation potentials) are calculated by
integration on a uniform spatial grid.

The ground state of the system is found using either exact
diagonalisation of the general eigenproblem (using \textsc{Scalapack}
routines) or a linear scaling approach\cite{Bowler:2002pt}.  For exact
diagonalisation, Bloch's theorem is applied and a Monkhorst-Pack mesh
of points in reciprocal space is generated.  For linear scaling, which
is purely real-space, this is not needed.
Forces\cite{Miyazaki:2004ee} and stresses are calculated as exact
derivatives of the energy (including Pulay forces and stresses from
the movement of the basis functions).

\subsection{Pseudopotentials}
\label{sec:pseudopotentials}

Over the last thirty years, pseudopotentials have become steadily more
accurate; during this period, various approaches have been developed
to soften and smooth the potentials, and reduce the necessary plane wave cutoff, including ultra-soft
pseudopotentials (USPP)\cite{Vanderbilt:1985gy} and the
projector-augmented wave (PAW)\cite{Blochl:1994rf}.  However, these
add complications to any implementation, and it is important to note
that Hamann's extension of norm-conserving
pseudopotentials\cite{Hamann:2013kg} to use multiple projectors for
each angular momentum (as suggested by
Vanderbilt\cite{Vanderbilt:1985gy}) permits extremely accurate
norm-conserving pseudopotentials to be developed.

Another important recent development in the area of pseudopotentials
is the ``delta'' study\cite{Lejaeghere:2014np,Lejaeghere:2016ls},
which compared the accuracy of all the various pseudopotential methods
to different all-electron approaches for a wide sampling of elemental
solids.  This study demonstrated that it is perfectly possible to
develop libraries of norm-conserving pseudopotentials that are as
accurate as the best PAW and USPP libraries.  To date, there are two sets of
these optimised norm-conserving (or ONCV) pseudopotentials\cite{Hamann:2013kg} freely available: the
PseudoDojo\cite{Setten:2018xl} and SG15\cite{Schlipf:2015bw}
libraries.  We have developed a PAO generation code for
\textsc{Conquest} which reads any ONCV pseudopotential generated by
Hamann's code, which includes both of these sets.  In this paper we
use the potentials from the PseudoDojo library, regenerated for these
tests using version 3.3.1 of Hamann's code, along with version 1.0.1
of the \textsc{Conquest} PAO generation code.

The only drawback to using these potentials is that almost all
elements have partial core corrections\cite{Louie:1982ax} included,
which can require a fine integration grid.  Moreover, many of the
heavier elements include semi-core states in the valence electrons,
requiring more bands to be solved.  These inclusions are necessary for
high accuracy; we intend to develop interfaces to other
pseudopotential generation schemes that will allow less accurate but
faster calculations, which will be reported in future work.

\section{Basis set sizes and defaults}
\label{sec:basis-set-sizes}

All basis sets have advantages and disadvantages; the well-known
advantages of plane waves (simplicity and systematic convergence)
apply for relatively small systems, where the efficiency of fast Fourier
transforms (FFTs) also applies; for large systems, however, the poor parallel
scaling of FFTs becomes significant.  Moreover, plane waves, being
solutions for the free electron, are poorly adapted to atomic
calculations (leading to the use of pseudopotentials) and more
generally to the solid state (leading to large numbers of basis functions).

There are approaches to systematic basis sets with localised orbitals,
where support functions (also known as non-orthogonal generalised
Wannier functions, among other names) are 
defined by a radius and a grid spacing equivalent to a plane wave
cutoff.  Examples of these include: the blip functions used in
\textsc{Conquest}\cite{Hernandez:1997ay}; periodic sinc
functions\cite{Mostofi:2002wb,Mostofi:2003ta};
wavelets\cite{Genovese:2008ya}; and Lagrange
functions\cite{Varga:2004xz}.  It is also possible to use the grid
points directly as basis functions with real space
techniques\cite{Beck:2000qy} including finite
elements\cite{Tsuchida:1996sp,Pask2005a} and finite
differences\cite{Fattebert:2006hf}.

There is, however, a simplicity and intuition which comes from using
atomic orbitals, or (in the case of a pseudopotential calculation)
pseudo-atomic orbitals, which consist of radial functions, normally
tabulated on a fine radial mesh, multiplied by appropriate spherical
harmonics.  Integrals between basis functions can be simplified, with
angular parts found analytically and radial grids so fine as to give
effectively analytic results.  The determination of the basis set then
relies on three questions: how are the radial functions calculated?;
how large an angular momentum is required?; and how many functions
should there be for each angular momentum?

The question of angular momentum is in large part determined by the
valence electrons being considered: these angular momenta must be
represented in the basis set.  It is then useful to consider
polarisation functions; following the Siesta code\cite{Soler:2002kn}
we can consider this as the effect of a local electric field on the
highest valence shell.  For elements up to the lanthanides, this
ensures that the PAO basis set includes angular momenta up to $l=2$;
unless an atom contains valence $f$ electrons, this is generally
sufficient.

The radial functions are always found with some form of confinement
(to ensure sparsity of matrices and efficiency).  While this
confinement can be a simple radial cutoff, it is not clear how to
choose that cutoff: a smaller cutoff will give greater efficiency, but
may not yield as good a basis function, particularly in materials where
long-range or weak interactions are key.  The Siesta code
introduced the idea of a uniform energy shift for all pseudo-atomic
orbitals\cite{Soler:2002kn} to allow for a consistent definition of
confinement for different angular momenta.  It also introduced a
flexible confinement potential to allow optimisation of the radial
functions\cite{Anglada:2002uq}.

Of course, with more than one radial function per angular momentum it
is possible to choose different radii, but the question of how to set
these is still difficult.  The \textsc{siesta} code uses a split norm idea for
subsequent radial functions: using a smooth polynomial from the origin
to a certain point, and beyond that point matching the original.
OpenMX calculates a set of five or six excited states (solutions with
increasing numbers of nodes) for each angular momentum in a hard
confining potential\cite{Ozaki:2003xq} and then combines these
primitive functions into a number of composite functions (typically
with two radial functions per valence state and one per polarisation
state), fitted to dimers or diatomic
molecules\cite{Ozaki:2004pt,Ozaki:2004qq}.  It is also possible to
optimise confining potentials by fitting to the eigenstates of
converged plane wave calculations\cite{Kenny:2000hq}.  As well as
pseudo-atomic orbitals, Gaussian orbitals are in use, for instance in
CP2K\cite{VandeVondele:2007sp}; Gaussian orbitals are the
primitive basis for the original method on which \textsc{Conquest}'s
multi-site support functions are based\cite{Rayson:2009vo}.

In developing PAO basis sets for \textsc{Conquest}, we have considered
the fact that the basis functions need to give variational flexibility
to the Kohn-Sham eigenstates, and therefore need to span a range of
radii.  The simplest way, then, is to start with the idea of energy
shifts and to specify radii on the basis of one energy that will give a
highly confined orbital and one that will give a loosely confined
orbital.  We have found, empirically, that energy shifts of 2\,eV and
0.02\,eV fulfil these criteria well, and have also found that for
situations where we specify a third radial function, using an
intermediate value of 0.2\,eV is effective.  In fact, for most systems
this gives three relatively evenly spaced confinement radii.

We define one set of PAOs by this equal energy criterion: a basis set
with one radial function uses 0.02\,eV; a set with two functions uses 2\,eV and 0.02\,eV; and
with three functions uses 2\,eV, 0.2\,eV and 0.02\,eV.  We then define four basis
sizes: minimal (one radial function, valence only or SZ); small (one
radial function for valence and polarisation, or SZP); medium (two
radial functions for valence and one for polarisation, or DZP); and
large (three radial functions for all angular momenta, or TZTP) though
in this paper we will not test the minimal basis.

We also consider an alternative, specifying \emph{equal radii} for all
angular momenta.  We do this by finding the radii for all angular
momenta for a given confinement energy, and taking their mean.  This gives
a uniform set of radii for all angular momenta.  This option tends to be more efficient, as
the overall radius of the largest support function is smaller than the
equal energy construction; however, it can require a finer integration
grid as the more diffuse functions are compressed.  We will see below
that the two approaches typically produce similar results, though the
equal radii approach is maybe a little better overall.

\section{Tests}
\label{sec:tests}

Our concern in this section is to test the accuracy of different PAO
basis sets against the converged plane wave result.  For this purpose
we use the PWSCF code from the QuantumEspresso suite\cite{Giannozzi:2009ox}, which reads
the same pseudopotentials as \textsc{Conquest} and allows a direct
comparison.  We note that we do not compare the results to experiments
as this is not our aim: this type of comparison has been performed
extensively elsewhere.

We have chosen various solid systems for our tests: three of the
elemental semiconductors (carbon, silicon and germanium, giving an
excellent sampling of different gap sizes); simple oxides (two
polymorphs of SiO$_{2}$ and MgO); perovskite oxides, including
materials with semi-core states (SrTiO$_{3}$, PbTiO$_{3}$ in the cubic
phase and the distorted perovskite MgSiO$_{3}$); an example of a
metallic system, non-magnetic bcc iron; and two weakly bonded systems
(the ordinary form of ice, ice XI, which has hydrogen bonds; and a
layered material, hexagonal boron nitride).  These systems represent
different forms of bonding and environment, and offer a good test of
the different basis sets sizes and truncation approaches.

\begin{table}
  \centering
  \begin{tabular}{lcccccc}
    & \multicolumn{2}{c}{Carbon}&\multicolumn{2}{c}{Silicon}&\multicolumn{2}{c}{Germanium}\\
    Basis & $a_{0}$(\AA) & $B_{0}$(GPa) & $a_{0}$(\AA) & $B_{0}$(GPa) & $a_{0}$(\AA) & $B_{0}$(GPa)\\
    PW & 3.558 & 449.3 & 5.431 & 93.28 & 5.676 & 67.47 \\
    \hline
    SZP(R) & 3.603 & 415.1 & 5.541 & 82.81 & 5.765 & 57.49\\
    DZP(R) & 3.573 & 444.1 & 5.458 & 90.96 & 5.692 & 68.70\\
    TZTP(R)& 3.562 & 452.9 & 5.437 & 91.70 & 5.690 & 64.99\\
    \hline
    SZP(E) & 3.600 & 425.4 & 5.533 & 81.20 & 5.741 & 59.73\\
    DZP(E) & 3.574 & 441.1 & 5.446 & 92.58 & 5.704 & 65.73\\
    TZTP(E)& 3.565 & 448.3 & 5.439 & 91.79 & 5.682 & 66.75\\
  \end{tabular}
  \caption{Structural parameters for C, Si and Ge.  Equal radii PAOs
    are indicated with (R), equal energy PAOs with (E).  All calculations
    used a reciprocal space mesh of $9\times 9\times 9$, a plane wave
    cutoff of 40Ha and an integration grid cutoff of 250Ha.}
  \label{tab:CSiGe}
\end{table}

The calculations used either the PBE functional\cite{Perdew:1996sj}
(Fe, SiO$_{2}$, MgO, MgSiO$_{3}$, ice XI and BN) or
the PBEsol functional\cite{Perdew:2008jk} (C, Si, Ge, SrTiO$_{3}$ and
PbTiO$_{3}$); the differences between the functionals
are relatively small, though PBEsol is considered generally better for
solids.  The same functional has been used for all
calculations of the same system (both QE and \textsc{Conquest}).  We
use the LibXC library\cite{Lehtola:2018td} for PAO generation and
\textsc{Conquest} calculations.

\textsc{Conquest} uses a regular grid for certain integrals, which is
set using an energy cutoff.  In these calculations the energy has been
converged with respect to this grid. In plane wave calculations the charge
density grid is generally taken to have an energy cutoff four times
that of the plane waves, and we note that the \textsc{Conquest} grid
energy cutoff is often around four times the QE plane wave cutoff.
Details of cutoffs and Brillouin zone sampling (using Monkhorst-Pack
grids) are given in the table captions.

\begin{figure}
  \centering
  \includegraphics[width=0.45\textwidth]{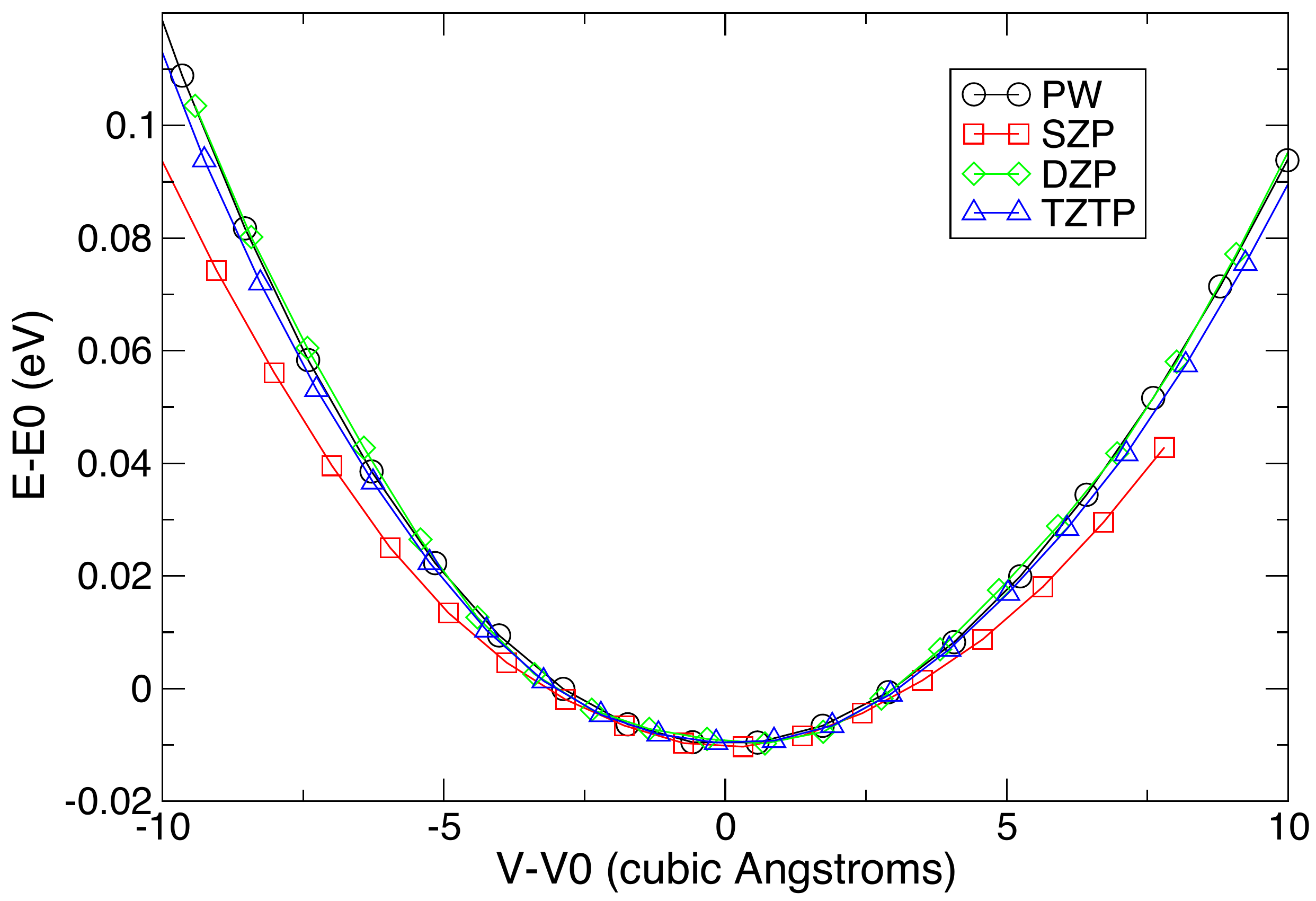}
  \caption{Binding energy curves for bulk Ge calculated with plane
    waves, and the three equal radii PAO basis sets.  Parameters as in the
    caption to Table~\protect\ref{tab:CSiGe}.}
  \label{fig:GeBEC}
\end{figure}

For the elemental semiconductors, shown in Table~\ref{tab:CSiGe}, we
see good agreement with DZP basis 
sets (differences up to 2-3\% of the bulk modulus and 0.5\% of the
lattice constant) and excellent agreement with TZTP basis sets (differences
typically less than 1\% of bulk modulus and 0.2\% of lattice
constant).  For context, when plotting binding energy curves (energy
versus volume), a difference of 1\% in the bulk modulus can barely be
distinguished (after correcting for differences in lattice
parameters).  We illustrate the worst agreement, for Ge, in
Fig.~\ref{fig:GeBEC} and the PAOs generated with equal radii (note
that all the curves are plotted relative to the fitted minimum volume
and energy).  In this
case, the TZTP basis set gives very slightly worse agreement than the DZP, which is
almost indistinguishable from the fully converged plane wave result.
The SZP is a little too soft, but is still respectable.  There is
little difference between the equal radii and equal energy basis sets,
except at SZP where the equal energy approach is maybe a little better.

\begin{table}
  \centering
  \begin{tabular}{lcccccc}
    & \multicolumn{2}{c}{$\alpha$-quartz}&\multicolumn{2}{c}{Stishovite}&\multicolumn{2}{c}{MgO}\\
    Basis & $V_{0}$(\AA$^{3}$) & $B_{0}$(GPa) & $V_{0}$(\AA$^{3}$) & $B_{0}$(GPa)& $V_{0}$(\AA$^{3}$) & $B_{0}$(GPa)\\
    PW     & 210.5 & 195.3 & 47.89 & 301.0 & 76.92 & 149.1\\
    \hline
    SZP(R) & 222.0 & 160.4 & 49.95 & 283.9 & 80.32 & 137.2\\
    DZP(R) & 215.6 & 177.1 & 49.16 & 289.2 & 78.49 & 141.5\\
    TZTP(R)& 213.0 & 193.5 & 48.25 & 295.0 & 78.51 & 148.3\\
    \hline
    SZP(E) & 220.8 & 165.9 & 49.82 & 260.4 & 80.31 & 149.7\\
    DZP(E) & 215.4 & 176.3 & 49.09 & 278.4 & 78.57 & 141.4\\
    TZTP(E)& 212.6 & 190.9 & 48.26 & 291.8 & 78.50 & 148.4\\
  \end{tabular}
  \caption{Parameters for SiO$_{2}$ in $\alpha$-quartz and stishovite
    structures.  Note that these are non-cubic, so the volume is
    given.  SiO$_{2}$ calculations used a plane wave cutoff of 40Ha
    and an integration grid cutoff of 200Ha with reciprocal space
    meshes of $3\times 2\times 3$ and $3\times 3\times 6$ for
    $\alpha$-quartz and stishovite, respectively.  MgO used a plane
    wave cutoff of 60Ha, an integration grid of 260Ha and a reciprocal
    space mesh of $4\times 4\times 4$.}
  \label{tab:SimpleOxides}
\end{table}

The simple oxides, with data found in Table~\ref{tab:SimpleOxides},
show similar performance to the elemental 
semiconductors.  The TZTP basis sets are converged to within 1\% of
the bulk modulus for the equal radii, but to within 2\% for the equal
energies, while the equilibrium volumes are all with 1\% (for both
$\alpha$-quartz and stishovite the unit cells are non-cubic, so we use
equilibrium volume rather than lattice constant).

A more challenging test for default PAOs (rather than PAOs tuned or
optimised to an environment) is to
compare the equilibrium phase (alpha quartz) to a high pressure
phase (stishovite).  The coordination of the Si atoms changes from
four to six between these two phases.  We compare the
energy difference per formula unit (i.e. per SiO$_{2}$ unit) for the
basis sets to the converged plane-wave result (-0.28 eV/unit).  The TZTP
basis sets give a good comparison (-0.20 eV/unit for equal energies and
-0.22 eV/unit for equal radii).  The DZP basis sets using perturbative
polarisation (the default setting) give the correct ordering (-0.08
eV/unit for equal energies, and -0.05 eV/unit for equal radii) if a
less accurate magnitude; however, generating $l=2$ orbitals simply as
excited eigenstates of the confined atom gives the incorrect ordering
for both DZP (+0.15 eV/unit for equal energies and +0.19 eV/unit for
equal radii) and SZP (+1.46 eV/unit for equal energies and +0.52
eV/unit for equal radii; these values are improved with perturbative
polarisation but still have the incorrect sign).  To test the
dependence further, we added a second polarisation
function (giving DZDP), which has an excellent result: -0.19 eV/unit for
equal energies, and -0.20 eV/unit for equal radii (for perturbative
polarisation functions; non-perturbative functions give a similar result).
Clearly comparisons of stability of structures require radial flexibility in
all angular momentum channels, and we would recommend use of at least
DZDP whenever considering this kind of problem with default (i.e. not
optimised) basis sets.  For structural properties of the
individual phases, however, the performance of
the DZP basis sets is reasonable, though not quite as accurate
as for the elemental semiconductors.

\begin{table}
  \centering
  \begin{tabular}{lcccccc}
    & \multicolumn{2}{c}{SrTiO$_{3}$}&\multicolumn{2}{c}{PbTiO$_{3}$}&\multicolumn{2}{c}{MgSiO$_{3}$}\\
    Basis & $V_{0}$(\AA$^{3}$) & $B_{0}$(GPa) & $V_{0}$(\AA$^{3}$) & $B_{0}$(GPa)& $V_{0}$(\AA$^{3}$) & $B_{0}$(GPa)\\
    PW     & 58.79 & 186.4 & 60.14 & 191.1 & 167.4 & 235.7 \\
    \hline
    SZP(R) & 60.99 & 170.0 & 61.66 & 189.2 & 170.5 & 198.2 \\
    DZP(R) & 60.15 & 180.7 & 60.69 & 190.9 & 168.0 & 223.7 \\
    TZTP(R)& 59.67 & 169.9 & 60.62 & 190.3 & 165.0 & 253.2 \\
    \hline 
    SZP(E) & 60.76 & 182.6 & 61.44 & 183.0 & 175.4 & 192.7 \\
    DZP(E) & 60.52 & 180.0 & 61.06 & 186.2 & 172.8 & 217.5 \\
    TZTP(E)& 60.08 & 183.4 & 60.83 & 187.8 & 169.7 & 246.1 \\
  \end{tabular}
  \caption{Parameters for cubic perovskites SrTiO$_{3}$ and
    PbTiO$_{3}$ and the distorted perovskite MgSiO$_{3}$.  STO and PTO
  used a plane wave cutoff of 40Ha, an integration grid cutoff of
  350Ha and a reciprocal space mesh of $9\times 9\times 9$.
  MgSiO$_{3}$ used a plane wave cutoff of 50Ha, an integration grid
  cutoff of 200Ha and a reciprocal space mesh of $3\times 3\times 2$.}
  \label{tab:Perovskites}
\end{table}

The perovskite structures, shown in Table~\ref{tab:Perovskites},
feature elements with semi-core states (4$s$ and 4$p$
states in Sr and 3$s$ and 3$p$ in Ti; the 5$d$ states for Pb might be
considered semi-core but are not in this case) which are described
with a single radial function.  MgSiO$_{3}$ combines the elements seen
in the simple oxides (we note that the pseudopotential in use here
includes the 2$s$ and 2$p$ states, and that they are treated as
semi-core).  The performance is excellent for PbTiO$_{3}$ 
and good for SrTiO$_{3}$ (though the bulk modulus for TZTP with equal
radii is surprisingly inaccurate) and the equilibrium volume for both
these materials is very close to the plane wave result.  It is
possible that treating the Sr 4$p$ states as valence, with further
radial functions, might improve the performance.  The bulk
modulus for MgSiO$_{3}$ is significantly worse than either MgO or
SiO$_{2}$, with errors of nearly 5\%.  Nevertheless, these results
give confidence in the default basis sets. 

\begin{table}
  \centering
  \begin{tabular}{lcc}
    Basis & $a_{0}$(\AA) & $B_{0}$(GPa) \\
    PW     & 2.758 & 271.4 \\
    \hline
    SZP(R) & 2.819 & 223.7 \\
    DZP(R) & 2.764 & 272.3 \\
    TZTP(R)& 2.758 & 276.5 \\
    \hline
    SZP(E) & 2.785 & 290.8 \\
    DZP(E) & 2.768 & 279.1 \\
    TZTP(E)& 2.769 & 272.2 \\
  \end{tabular}
  \caption{Parameters for non-magnetic bcc Fe.  Calculations used a
    plane wave cutoff of 50Ha, an integration grid cutoff of 200Ha and
  a reciprocal space mesh of $5\times 5\times 5$.}
  \label{tab:BCCFe}
\end{table}

Metallic bonding is very different to the covalent and ionic bonding
studied thus far, but the performance of the default basis sets for
non-magnetic bcc Fe, shown in Table~\ref{tab:BCCFe}, is
excellent (we chose non-magnetic Fe simply for convenience;
  \textsc{Conquest} is capable of spin-polarised operation as simply
  as non-spin-polarised).  In this case, both TZTP basis sets reproduce the plane
wave results (the equal energy case has errors of 0.5\% while the
equal radii 2\% in bulk modulus) while the DZP give excellent
results.  The equal energy SZP is still reasonable, though the equal
radii is a little inaccurate.  (For the Fe atom, both 3$s$ and 3$p$ states are included
as semi-core states.)

\begin{table}
  \centering
  \begin{tabular}{lccc}
    Basis   & a (Bohr) & b  (Bohr) & c  (Bohr)\\
    PW      & 8.20 & 14.41 & 13.36 \\
    \hline
    SZP(R)  & 7.84 & 13.62 & 12.75 \\
    DZP(R)  & 7.85 & 13.87 & 13.02 \\
    TZTP(R) & 8.30 & 14.28 & 13.38 \\
    \hline
    SZP(E)  & 7.76 & 13.43 & 12.69 \\
    DZP(E)  & 7.82 & 13.79 & 12.74 \\
    TZTP(E) & 8.29 & 14.29 & 13.44
  \end{tabular}
  \caption{Ice XI relaxed cell parameters (using PBE).  Calculations
    used a plane wave cutoff of 40Ha, an integration grid cutoff of
    150Ha and a reciprocal space mesh of $3\times 2\times 2$.}
  \label{tab:ice}
\end{table}

Weakly bound systems offer a larger challenge to local orbital basis
sets (putting to one side the issues that DFT has with these systems,
which will be the same for any other basis set).  We start with
hydrogen bonding, considering the optimum unit cell for ice XI, the
ordinary form of ice, shown in Table~\ref{tab:ice}.  With a TZTP basis
set, the three parameters are all accurate, comparing to the PW result to better than 1\%, while
DZP and SZP are much less accurate.  This is the first case we have
found where the use of a default (non-optimised) DZP basis set might cause a
significant error.  It is interesting to see how effective the TZTP
basis sets are, even for a difficult system like this.

\begin{table}
  \centering
  \begin{tabular}{llcc}
    Functional & Basis  & Distance (Bohr) & Energy (meV/atom) \\
    \hline
    PBE        & PW     & 7.86     & 0.99   \\
    \hline
               & SZP(R) & 6.62     & 6.25   \\
               & DZP(R) & 7.96     & 0.98   \\
               & TZTP(R)& 7.90     & 1.64   \\
    \hline
               & SZP(E) & 6.66     & 8.79   \\
               & DZP(E) & 7.58     & 4.23   \\
               & TZTP(E)& 7.95     & 1.45   \\
    \hline
    PBE-D2     & PW     & 5.84     & 31.89  \\
    \hline
               & SZP(R) & 5.72     & 43.20  \\
               & DZP(R) & 5.99     & 28.59  \\
               & TZTP(R)& 6.03     & 28.45  \\
    \hline
               & SZP(E) & 5.75     & 44.82  \\
               & DZP(E) & 6.02     & 28.73  \\
               & TZTP(E)& 6.02     & 28.27  \\
  \end{tabular}
  \caption{BN optimum inter-layer distances and minimum energies for PBE and
    dispersion corrected (DFT-D2) PBE.  Calculations
    used a plane wave cutoff of 50Ha, an integration grid cutoff of
    150Ha and a reciprocal space mesh of $3\times 2\times 1$.
    \textsc{Conquest} calculations used a counterpoise correction\cite{Duijneveldt:1994no}}
  \label{tab:BNweak}
\end{table}

Finally we turn to the layered material, boron nitride, which has
dispersion interactions between layers, where the range of the PAOs may play a key
role in the DFT part of the interaction; for the \textsc{Conquest} calculations,
counterpoise corrections were used\cite{Duijneveldt:1994no} to account
for basis-set superposition errors.  Data for the minimum distance
between layers, and the resulting interaction energy is shown in
Table~\ref{tab:BNweak}.  We report results for both standard PBE and
PBE with semi-empirical dispersion corrections\cite{Grimme:2006kq};
again, we note that we are testing the accuracy of our default PAO
basis sets against converged PW calculations (and not the accuracy of DFT).

Good accuracy in comparison to the PW results is achieved with DZP and TZTP for the equal radii and
equal energy basis sets for both approaches, though the equal energy
DZP basis is perhaps a little high in the PBE interaction energy.  The
SZP basis set is less accurate, though it is interesting that the
inter-layer spacing is rather close with the PBE-D2 approach,
suggesting that the dispersion corrections mask the basis set errors.

\section{Conclusions}
\label{sec:conclusions}

We have introduced an approach to the construction of PAO basis sets
for the \textsc{Conquest} code based on increasing energies (spaced by
factors of ten).  The radial functions are either based on a geometric
increase in energy (equal energies) or on equal radii based on the
average of the energies for all angular momenta (equal radii).  These
are both implemented in the \textsc{Conquest} PAO generation code
which will be made available with the \textsc{Conquest} code on its
release.

We have compared the performance of small (SZP), medium (DZP) and
large (TZTP) basis sets with exact diagonalisation to converged plane
wave results using the same functionals and pseudopotentials.
Overall, the large (TZTP) basis sets are shown to reproduce the plane wave
results to better than 1\% in bulk modulus and often within 0.1\% of
the lattice constant for a variety of solids.  Except for weakly
bonded systems (particularly ice) we found that the medium (DZP)
basis sets are nearly as accurate as the large basis sets and will
typically be 5--10 times faster (simply from the reduction in matrix
size).  In general, both approaches (equal energies and equal radii)
produce reliable results, though the equal radii approach gives
slightly better results overall, and with smaller overall radii will give
better efficiency.  Alongside results from codes such as
FHI-aims\cite{Blum:2009uk}, these results demonstrate that local orbital codes
can be as accurate as converged plane wave codes.

\ack
This work was supported by World Premier International Research Center
Initiative (WPI Initiative) on Materials Nanoarchitectonics (MANA),
New Energy and Industrial Technology Development Organization of Japan
(NEDO) Grant (P16010), “Exploratory Challenge on Post-K computer” by
MEXT, and JSPS Grant-in-Aid for Scientific Research (18H01143,
17H05224).  The authors are grateful for computational support from the UK
Materials and Molecular Modelling Hub, which is partially funded by
EPSRC (EP/P020194), for which access was obtained via the UKCP
consortium and funded by EPSRC grant ref EP/P022561/1.  They also
acknowledge computational support from the UK
national high-performance computing service, ARCHER, for which access
was obtained via the UKCP consortium and funded by EPSRC grant ref
EP/K013564/1.

\newcommand{\newblock}{}
\bibliographystyle{jpcm}
\bibliography{paos}

\end{document}